\documentclass[epj]{article}
\usepackage{blindtext}
\usepackage{graphics}
\usepackage{amsmath,amssymb,graphicx}
\usepackage{graphicx}
\usepackage{booktabs}
\usepackage[colorlinks=true, allcolors=blue]{hyperref}
\usepackage{epsfig}

\newcommand\nonu{\nonumber}
\newcommand\br{\begin{eqnarray}}
\newcommand\er{\end{eqnarray}}
\newcommand\brs{\begin{eqnarray*}}
\newcommand\ers{\end{eqnarray*}}
\newcommand\be{\begin{equation}}
\newcommand\ee{\end{equation}}

\title{ Higher order Soliton Complexes   in Coupled Nonlinear Schr\"odinger  Equation with 
Variable Coefficients}
\author{Sudipta Nandy\thanks{sudiptanandy@gmail.com} \  \ and  Abhijit Barthakur 
\\ 
Cotton University,  Panbazar, Guwahati -781001, India}
\date{\today}
\begin{document}
\maketitle
\begin{abstract}
We present the explicit dark-bright three soliton solution  and the associated spectral problem for the  variable 
coefficient integrable coupled NLS equation.  
Using asymptotic analysis as well as graphical analysis we study the interactions in 
soliton complexes. We present a correlation  between the soliton parameters and the interaction pattern in 
three soliton complexes. Using asymptotic analysis, we present a few 
interesting features of  complex three soliton bound state and  interaction of  dark-bright two soliton complex 
with a regular soliton. Using three soliton interactions we have shown that  
the energy sharing take place between soliton even when the soliton do not collide with each other. 
The results found by us might be useful for the development of soliton control, all optical gates 
as well as all optical switching devices. We hope that  the analysis of three soliton complexes would be useful for a 
better understanding of soliton interactions in nonlinear fiber  as well as in a bulk medium.
\end{abstract}
{\small 05.45.Yv,  42.81Dp, 42.65.Tg}

\section{Introduction}\label{Introd}
Since it's discovery in 1965 by Kruskal \cite{KRUS65}, the potential  application of soliton has been explored  in 
many areas of physical sciences, such as in fluid mechanics and  plasma physics\cite{KUZNETSOV86}, in cold 
atoms \cite{{KAVREKIDIS09},{DALFOVO99}}, in high energy physics\cite{GP97} etc. Among the integrable systems
perhaps nonlinear Schr-\"odinger equation (NLS) has the widest applications. Optical soliton is one such 
application. In early seventies of twentieth century 
Hasegawa through a scientific communication \cite{HASEGAWA73}, explained  that there exist an entity called
optical soliton which maintains a balance  between dispersion and  nonlinearity as it propagates through a dispersive
dielectric medium and the dynamics of the optical soliton can be described by NLS equation. 
A uniform  optical medium with anomalous dispersion supports a bright soliton whereas  a dark soliton(described as  
a dip in a continuous background) is supported by a medium with normal dispersion. Incidentally dark solitons are 
more stable in noisy background  and under perturbations and  they  interact  weakly compared to bright 
solitons\cite{Foursa96}. 
Over the past three decades there have been remarkable advances in both theoretical 
and experimental research on soliton and its applications \cite{{GPA95},{KIV03}} (and the references therein).  
A dispersion managed(DM) soliton\cite{GPA95} however,  is a more advanced concept, where solitons may exist in a 
normal and anomalous dispersion coexisting media, and  is described by the standard nonlinear 
Schr\"odinger equation model with varying dispersion and nonlinear coefficients and gain\cite{Serkin00}. 
A DM soliton can be accelerated or retarded and 
amplified while preserving  its shape. Recently there have been many important publications on the dynamics 
of a bright soliton \cite{{XG09},{ZY11},{Ganapathy08}} and dark soliton 
\cite{{MPNSD17},{LIU13},{SERKIN10},{KIVSHAR98},{Sheppard97}} based on the model 
given in \cite{Serkin00} and  a more recently developed non-autonomous soliton model\cite{Serkin02}, where 
the authors predicted various applications of DM soliton. 

In comparison to scalar soliton a vector soliton (with more than one mutually self-trap components), 
proposed by Manakov \cite{SVM74} demonstrates interesting additional features. For instance, bright-bright soliton 
pair (bright soliton 
in all the modes), bright-dark soliton pair (modes are shared between bright solitons and dark solitons) 
\cite{{SERKIN93},{YAN12},{Radhakrishnan959707}},  dark-dark  soliton pair (dark solitons in all the modes) 
\cite{OHTA11} obtained in a normal dispersion and anomalous dispersion
coexisting  dielectric medium,  soliton shape changing  in an inelastic soliton interaction 
\cite{Radhakrishnan97}, simulation of electronic logic gates using elastic and inelastic soliton 
interactions \cite{{STEIGLITZ2009},{STEIGLITZ2000},{MJ2004}}. All these discoveries has made the field of applications 
of vector solitons more exciting and vibrant.

It is now natural to ask whether a DM vector soliton is also possible and if possible what would be the dynamics, how 
do they interact and their possible applications.  Secondly whether the mathematical techniques available for the scalar soliton 
and vector solitons are applicable to 
DM vector soliton also. In the literature other than few  notable publications  
\cite{{Zhong10},{Wen12},{HAN2014},{KRAENKEL2016},{CHOW2011}}, study in this direction is 
comparatively sparse. Present authors in one of their recent works  investigated the two soliton interactions using the  
variable coefficient  NLS model\cite{SN15}. 

A three soliton solution while being an important signature of integrability \cite{JH88},  shows many 
interesting features.  An interesting aspect of  
three soliton solution is the bound state soliton interactions, many of which are unlike normal soliton interactions, as found in
two soliton bound state interactions\cite{{YAN12},{SN15}}.  A  three soliton bound state  where a two soliton  
complex  interact with a third soliton, is expected to bring forth more interesting facts about the soliton interaction. 
Keeping in context the above facts, here we shall 
investigate three soliton solution of  a Lax integrable variable coefficient 2-coupled NLS model.
 We shall also  analyze  interactions in dark-bright soliton complexes
using asymptotic analysis, wherever applicable and 
the remaining cases using graphical analysis. Such analysis might contain useful information that would be necessary 
for further development in the applications of optical soliton. 

The organization of the paper is as follows. Section \ref{Introd} contains introduction. 
In section \ref{Model}  the model for the proposed study is described and the associated Lax pair  is presented. 
Bilinear form and  soliton solutions of the proposed model  is also included in this section. 
In section \ref{Asymp} soliton interactions are studied using asymptotic analysis.
In section \ref{complex} three soliton complexes are studied using asymptotic as well as graphical analysis. 
Section \ref{Conclusion} is the concluding one.

\section{THE MODEL EQUATION}\label{Model}
We consider the  following generalized NLS equation with variable coefficients\cite{SN15}:
\begin{equation}
\label{CDM1}
{\bf i} {\bf q_j}_t + \frac{D(t)}{2} {\bf q_j}_{zz} +  \gamma R(t)\displaystyle{\sum_{l=1}^{n}}\sigma_l
|{\bf q}_l|^2 {\bf q_j} = {\bf i} \Gamma(t){\bf q_j}
\end{equation} 
where $ { q_j}$ ( $ j=1, \cdots n$) are complex amplitude of the $j^{th}$ field component, 
of an inhomogeneous dispersive and nonlinear medium, subscript $t$ and $z$ are the dimensionless parameters, 
and denote the partial derivatives   with respect to $t$ and $z$ respectively,
$ D(t) $  and  $R(t)$ 
denote the variable dispersion 
coefficient  and nonlinearity coefficient respectively. $\Gamma(t)= \frac{\partial_t D(t) R(t)-D(t) 
\partial_t R(t)}{2 R(t) D(t)}$ and $\sigma_l$ ($=\pm 1$) define the sign of 
the nonlinearity.  $\sigma_l= -1$(+1) stands for a defocusing (focusing) type nonlinearity. 
If $\sigma_l = +1$ (for $l=1,2,\cdots n$) nonlinearity is only focusing type and  if 
$\sigma_l = -1$ (for $l=1,2,\cdots n$) nonlinearity is defocusing type. 
If $\sigma_l = +1$ (for $l=1,2,\cdots k$) and $\sigma_l =-1$ (for  $l=k+1,k+2,\cdots n$) then both 
focusing (for $k$ components) and defocusing (for $(n-k)$ components) type nonlinearity occur at the same time.\\
Bright-dark type soliton solution for eq.\ref{CDM1}, with bright solitons in $k$ modes and dark solitons 
in $ (n-k) $  modes  can be obtained with  ($ 2^n -1$) possible combinations of $\sigma_l = \pm 1$ \cite{SN15}. 
Notice that the choice $ D(t) =R(t) =1$ and $\sigma_l= 1$ lead eq. \ref{CDM1} to standard Manakov model
\cite{SVM74}.   
    
In the present paper  we shall obtain dark-bright type $3$-soliton for eq.\ref{CDM1} using the 
Hirota \cite{Hirota8074} method.  
Subsequently using asymptotic and graphical analysis we present a correlation between soliton parameters and the 
pattern of interaction in a  3-soliton complex.
The  analysis here can  be generalized to  system with ($k>2$) modes  for bright soliton  and (($n-k>1$)) modes for dark 
soliton. 
However, it is worthwhile to mention here that  in  Manakov system, adding   
number of  modes do not add more intricacies in to the system, as pairwise collision in N($>2$)-coupled(NLS 
equation can be reduced to pairwise collision in 2-coupled NLS equation \cite{STEIGLITZ2003}.
  
Secondly, using asymptotic analysis we shall explicitly show that there is never an energy sharing between the dark 
solitons during an interaction. Thus in the proposed study, we consider  a minimum, yet sufficient number of modes, 
in NLS eq.\ref{CDM1}.

\subsection{Associated spectral problem }\label{Spectral}
The Lax pair associated with the eq. \ref{CDM1} is:
\br
\begin{aligned}
\partial_z {\bf \Psi(z,t)} &=& {\bf U}{\bf \Psi(z,t)}\\
\partial_t {\bf \Psi(z,t)} &=& {\bf V}{\bf \Psi(z,t)}
\end{aligned}
\er
where ${\bf U}$ and ${\bf V}$ respectively are $ (n+1) \times (n+1) $ matrices.
The explicit form of  ${\bf U} $ and ${\bf V}$ are:

\br
\begin{aligned}
{\bf U} &= - {\bf i} \lambda {\bf \Sigma}  +  \sqrt{\frac{R(t)}{D(t)}}{\bf A}\qquad \qquad \qquad \qquad\qquad \qquad \qquad \qquad \qquad \qquad \qquad  \qquad\\
{\bf V} &= -{\bf i}{\bf \Sigma}\{ \frac{{\bf R(t)}}{2} {\bf A}^2  + \int^t \lambda_t ds +
D(t)\lambda^2 -\gamma_0(t) - \sqrt{\frac{R(t)}{D(t)}} \frac{D(t)}{2} {\bf A}_z \} \\
& + \sqrt{\frac{R(t)}{D(t)}} D(t) \lambda {\bf A}
\end{aligned}
\er

\br
\begin{aligned}
{\bf \Sigma} &= \left(\begin{array}{cccc}
          1 & 0 & \cdots & 0 \\
          0 & \ddots & \vdots & \vdots  \\
          \vdots & \cdots & 1 & 0 \\
          0 & \cdots & 0  & -1
          \end{array} \right)_{(n+1) \times (n+1)}
\end{aligned}
\er
\br
\begin{aligned}
{\bf A} &= \left(\begin{array}{ccccc}
          0 & \cdots & \cdots & 0 & {\bf \sqrt{\sigma_1} q}_1\\
          \vdots & \ddots & ~ & \vdots &{\bf \sqrt{\sigma_2} q}_2 \\
          \vdots & ~ & \ddots &\vdots &\vdots \\
          0 & \cdots & \cdots & 0 & {\bf \sqrt{\sigma_n} q}_n\\
          -{\bf \sqrt{\sigma_1} q}_1^* & -{\bf \sqrt{\sigma_2} q}_2^* & \cdots & -{\bf \sqrt{\sigma_n} q}_n^* & 0
          \end{array} \right)_{(n+1) \times (n+1)} 
\end{aligned}
\er
where  $ \lambda $ is the spectral parameter. The compatibility condition namely, $ {\bf U}_t - {\bf V}_z + UV-VU=0$ 
gives the eq. \ref{CDM1} provided $\int_0^t \lambda_t ds = \gamma_0(t)$.

\subsection{Bilinearization and Soliton Solutions using Hirota's Method}\label{Hirota}
 
In order to write eq.\ref{CDM1} in the bilinear form, we make the following bilinear transformation, 
\br
\begin{aligned}
\label{BL1}
q_j &= \frac{g^{(j)}(t,z)}{f(t,z)};  \text{(bright soliton)}\\
q_l &= \frac{g^{(l)}(t,z)}{f(t,z)};  \text{(dark soliton)}
\end{aligned}
\er
where $g^{(j)}(t,z)$ and $g^{(l)}(t,z)$ are complex and $f(t,z)$ is real. Consequently in the new set of 
variables we have the following set of bilinear equations:
\br
\label{BL2}
\begin{aligned}
({\bf i} D_t + \frac{D(t)}{2} D_z^2 -\lambda)(g^{(j)} . f) 
&= {\bf i} \Gamma(t)(g^{(j)}.f) \\
(\frac{D(t)}{2} D_z^2 - \lambda)(f.f) 
&= \gamma R(t) \sum_{l=1}^n \sigma_l g^{(l)}{g^{(l)*}} 
\end{aligned}
\er
which follows from eq.\ref{CDM1}. $D_t$ and $D_z^2$ are Hirota derivatives \cite{{Hirota8074},{SN10}} and are defined by, 
\br
D^m_z D_t^n u(z,t)v(z,t)&=&(\frac{\partial}{\partial z} - \frac{\partial}{\partial z^{\prime}})^m
(\frac{\partial}{\partial t} - \frac{\partial}{\partial t^{\prime} })^n \nonumber \\
&\times& u(z,t)v(z^{\prime},t^{\prime})|_{z^{\prime}=z;\quad t^{\prime}=t} \nonumber 
\er
In order to obtain the soliton solutions $ g^{(j)} $ ($j=1,2,\cdots k$), $ g^{(l)} $ 
($l= k+1, k+2, \cdots n$) and $f$ are expanded with 
respect to an arbitrary parameter $\epsilon$ as follows,

\begin{eqnarray}
\label{HEXP}
\begin{aligned}
g^{(j)}&= \epsilon g_1^{(j)} + \epsilon^3 g_3^{(j)}+ \cdots  \\
g^{(l)}&= g_0^{(l)}( 1 + \epsilon^2 g_2^{(l)}  + \cdots ) \\
f &= 1 + \epsilon^2 f_2 +  \cdots  
\end{aligned}
\end{eqnarray}

Dark-Bright  one soliton, two soliton, and three soliton solutions  are obtained from the following expressions,
\br
\begin{aligned}
q_{j1} &= \frac{\epsilon g_1^{(j)}}{1 + \epsilon^2 f_2 }|_{\epsilon=1}; \quad \text{bright soliton}\\
q_{l1} &= \frac{g_0^{(l)}( 1 + \epsilon^2 g_2^{(l)} ) }{1 + \epsilon^2 f_2  } |_{\epsilon=1};  
\quad \text{dark soliton}
\end{aligned}
\er

\br
\begin{aligned}
q_{j2} &= \frac{\epsilon g_1^{(j)}  +\epsilon^3 g_3^{(j)} }{1 + \epsilon^2 f_2  +\epsilon^4 f_4 }|_{\epsilon=1}; 
 \quad  \text{bright 2-soliton} \\
q_{l2} &= \frac{g_0^{(l)} ( 1 + \epsilon^2 g_2^{(l)}+ \epsilon^4 g_4^{(l)} ) }
{1 + \epsilon^2 f_2  + \epsilon^4 f_4 } |_{\epsilon=1};    \quad \text{dark 2-soliton}
\end{aligned}
\er

\br
\begin{aligned}
\label{3SS}
q_{j3} &= \frac{\epsilon g_1^{(j)}  +\epsilon^3 g_3^{(j)} +\epsilon^5 g_5^{(j)}}
{1 + \epsilon^2 f_2  + \epsilon^4 f_4  + \epsilon^6 f_6 }|_{\epsilon=1}; \\
&~  \text{bright 3-soliton} \\
q_{l3} &= \frac{g_0^{(l)} ( 1 + \epsilon^2 g_2^{(l)}+ \epsilon^4 g_4^{(l)} +\epsilon^6 g_6^{(l)} ) }
{1 + \epsilon^2 f_2  + \epsilon^4 f_4  + \epsilon^6 f_6 }|_{\epsilon=1}; \\
 &~  \text{dark 3-soliton}
\end{aligned}
\er
(for $j=1,2,\cdots,k$ and $l= k+1, k+2, \cdots n$)\\
Explicit one soliton and two soliton solutions of eq.\ref{CDM1} are obtained  in \cite{SN15}.
However,  three soliton solution is one of the important criteria  for the existence of $N$-soliton 
solutions \cite{JH88}.  Substituting eq.\ref{3SS} in eq.\ref{BL2} we obtain the three soliton solution
of eq.\ref{CDM1}. 
In this paper we consider  $j=1,2$ and $l=3$, that is  bright 3-soliton $3SS$ is in component $ 1, 2 $ and 
dark $3SS$  is in component  $3$.  Explicit solution is given in the Appendix.

\section{Soliton Interaction and Asymptotic Analysis } \label{Asymp}

Energy sharing in Manakov solitons during an interaction is an well established fact. In the present system however, 
the  interaction energy sharing is a  periodic phenomena. 
Let us analyze the asymptotic behavior of the $3$-soliton 
solution eq.\ref{3SS} (see Appendix ), where a bright three soliton is in two components  and a 
dark three soliton is in one component in a three components system. 
 Asymptotically  solitons, irrespective of their coordinates
are sufficiently separated,  such that there is no interaction among them. 
Consider for instance, at an instant   when soliton  $q_3^{(j)}$  is at $ z \rightarrow  \infty $,
$q_1^{(j)}$  and  $q_2^{(j)}$ are  at  $ z \rightarrow  -\infty $  and they are moving with  relative velocities as 
they are approaching each other. Note that the sign before $ z$ are interchangeable as the solitons change their positions
periodically.  Then the asymptotic  expressions of bright soliton are,  as $ z \rightarrow -\infty $ 
\br
\begin{aligned}
q_1^{(j)}\vert^{-} & =   \frac{\alpha_1^{(j)} (t)}{\sqrt{\delta_1 }}
Sech(\frac{\theta_1 + \theta_1^* +ln(\delta_1)}{2}) e^{\frac{\theta_1-\theta_1^*}{2} }; \quad
\theta_2  + \theta_2^*+\theta_3 + \theta_3^* \rightarrow -\infty \\
q_2^{(j)}\vert^{-} & =   \frac{\alpha_2^{(j)} (t)}{\sqrt{\delta_2 }}
Sech(\frac{\theta_2 + \theta_2^* +ln(\delta_5)}{2}) e^{\frac{\theta_2-\theta_2^*}{2} }; \quad
\theta_1 + \theta_1^*+\theta_3 + \theta_3^* \rightarrow -\infty \\
q_3^{(j)}\vert^{-} & =   \frac{\varsigma_{1}^{(j)}}{\sqrt{\Upsilon \varrho_1 }}
Sech(\frac{\theta_3 +  \theta_3^* + ln(\frac{\Upsilon}{\varrho_1})}{2}) e^{\frac{\theta_3-\theta_3^*}{2 }}; \quad
\theta_1  + \theta_1^*+\theta_2 + \theta_2^* \rightarrow  + \infty 
\end{aligned}
\er

and as $ z \rightarrow +\infty $
\br
\begin{aligned}
 q_1^{(j)}\vert^{+} & = \frac{\varsigma_{3}^{(j)}}{\sqrt{ \Upsilon \varrho_9 }}
Sech(\frac{\theta_1 +  \theta_1^* + ln( \frac{\Upsilon}{\varrho_9})}{2})e^{\frac{\theta_1-\theta_1^*}{2 }};\quad
\theta_2 + \theta_2^*+\theta_3 + \theta_3^* \rightarrow + \infty  \\
q_2^{(j)}\vert^{+} & = \frac{\varsigma_{2}^{(j)}}{\sqrt{ \Upsilon \varrho_5 }}
Sech(\frac{\theta_3 +  \theta_3^* + ln( \frac{\Upsilon}{\varrho_5})}{2}) e^{\frac{\theta_2-\theta_2^*}{2 }}; \quad
 \theta_1 + \theta_1^*+\theta_3 + \theta_3^* \rightarrow + \infty \\
q_3^{(j)}\vert^{+} & =  \frac{\alpha_3^{(j)}(t) }{\sqrt{\delta_9}}
Sech(\frac{\theta_3 + \theta_3^* + ln(\delta_9)}{2}) e^{\frac{\theta_3 -\theta_3^*}{2}}; \quad
\theta_1  + \theta_1^*+\theta_2 + \theta_2^* \rightarrow -\infty
\end{aligned}
\er

Similarly the asymptotic expression of dark solitons are, 
as  $ z  \rightarrow -\infty $ 

\br
\begin{aligned}
\label{ASD1}
q_1^{(3)}\vert^{-} & =  \chi^{(3)}(t)
( 1 + \frac{\gamma_1^{(3)}}{\delta_1} + (\frac{\gamma_1^{(3)}}{\delta_1} -1) \\ 
&\times 
Tanh( \frac{\theta_1 + \theta_1^* +ln({\delta_1}) }{2}) e^{\phi^{(3)} + \frac{\theta_1 - \theta_1^*}{2}}; \quad
\theta_2  + \theta_2^*+\theta_3 + \theta_3^* \rightarrow - \infty \\
q_2^{(3)}\vert^{ - } & =  \chi^{(3)}(t)
(1 + \frac{\gamma_5^{(3)}}{\delta_5} + (\frac{\gamma_5^{(3)}}{\delta_5} -1) \\ 
&\times  
Tanh( \frac{\theta_2 + \theta_2^* +ln({\delta_5}) }{2})e^{\phi^{(3)} + \frac{\theta_2 - \theta_2^*}{2}}; \quad
\theta_1  + \theta_1^*+\theta_3 + \theta_3^* \rightarrow - \infty \\
q_3^{(3)}\vert^{-} & =  \chi^{(3)}(t) \frac{\rho_1^{(3)}}{ \varrho_1}
( 1 + \frac{\gamma_9^{(3)}}{\delta_9} + (\frac{\gamma_9^{(3)}}{\delta_9} -1) \\ 
&\times 
Tanh(\frac{\theta_3 + \theta_3^* + ln(\frac{\Upsilon}{\varrho_1})}{2} )e^{\phi^{(3)} 
+ \frac{\theta_3 - \theta_3^*}{2}}; \quad
\theta_1 + \theta_1^* +\theta_2 + \theta_2^* \rightarrow + \infty
\end{aligned}
\er

and as  $z  \rightarrow +\infty$  \\
\br
\begin{aligned}
\label{ASD2}
q_1^{(3)}\vert^{ + } & =  \chi^{(3)} (t)\frac{\rho_9^{(3)}}{ \varrho_9}
( 1 + \frac{\gamma_1^{(3)}}{\delta_1} + (\frac{\gamma_1^{(3)}}{\delta_1} -1)\\  
&\times Tanh( \frac{\theta_1 + \theta_1^* +ln({\frac{\Upsilon}\varrho_9})}{2})e^{\phi^{(3)} + \frac{\theta_1 - \theta_1^*}{2}};\quad
\theta_2 + \theta_2^*  + \theta_3 + \theta_3^*  \rightarrow + \infty \\
q_2^{(3)}\vert^{ + } & =  \chi^{(3)} (t)\frac{\rho_5^{(3)}}{ \varrho_5}
( 1 + \frac{\gamma_5^{(3)}}{\delta_5} + (\frac{\gamma_5^{(3)}}{\delta_5} -1)  \\
&\times Tanh( \frac{\theta_2 + \theta_2^* +ln({\frac{\Upsilon}\varrho_5})}{2})e^{\phi^{(3)} + \frac{\theta_2 - \theta_2^*}{2}};\quad
\theta_1 + \theta_1^*  + \theta_3 + \theta_3^*  \rightarrow + \infty \\
q_3^{(3)}\vert^{-} & =  \chi^{(3)} (t)
 ( 1 + \frac{\gamma_9^{(3)}}{\delta_9} + (\frac{\gamma_9^{(3)}}{\delta_9} -1)\\ 
&\times Tanh(\frac{\theta_3 + \theta_3^* + ln(\delta_9)}{2} )e^{\phi^{(3)} + \frac{\theta_3 - \theta_3^*}{2}};\quad
\theta_1 + \theta_1^* + \theta_2 + \theta_2^* \rightarrow - \infty
\end{aligned}
\er

\noindent
Let  ${\cal A}_j^{1-}$, ${\cal A}_j^{2-}$ ,  ${\cal A}_j^{3-}$ denote the amplitudes and 
$\Phi_j^{1-}$,  $\Phi_j^{2-}$  $\Phi_j^{3-}$denote the phase of 
soliton $1$, soliton $2$   and  soliton $3$ respectively before interaction  and ${\cal A}_j^{1+}$, ${\cal A}_j^{2+}$ and 
${\cal A}_j^{3+}$ denote the 
amplitudes and $\Phi_j^{1+}$,  $\Phi_j^{2+}$ and $\Phi_j^{3+}$ denote the phase of the three 
solitons after interaction. Then under one of the following conditions, 
\be
\label{EL1}
\frac{\alpha_1^{(1)}(t)}{\alpha_1^{(2)}(t)} = \frac{\alpha_2^{(1)}(t)}{\alpha_2^{(2)}(t)}
=\frac{\alpha_3^{(1)}(t)}{\alpha_3^{(2)}(t)}
\ee

\br
\label{EL2a}
\begin{aligned}
\alpha_1^{(1)}(t) & = \alpha_2^{(2)}(t) = \alpha_3^{(2)}(t)=0\\
 \alpha_2^{(1)}(t) &\ne 0;  \alpha_1^{(2)}(t) \ne 0; \alpha_3^{(1)}(t) \ne 0 
\end{aligned}
\er
\br
\label{EL2b}
\begin{aligned}
\alpha_2^{(1)}(t) & = \alpha_1^{(2)}(t) = \alpha_3^{(2)}(t)=0; \\
\alpha_1^{(1)}(t) &\ne 0; \alpha_2^{(2)}(t)\ne 0;   \alpha_3^{(1)}(t) \ne 0 
\end{aligned}
\er

\br
\label{EL2bc}
\begin{aligned}
\alpha_1^{(2)}(t) & = \alpha_2^{(2)}(t) = \alpha_3^{(1)}(t)=0; \\
\alpha_1^{(1)}(t) &\ne 0; \alpha_2^{(1)}(t)\ne 0;   \alpha_3^{(2)}(t) \ne 0 
\end{aligned}
\er

\begin{figure}
  \includegraphics[width=1.0\columnwidth]{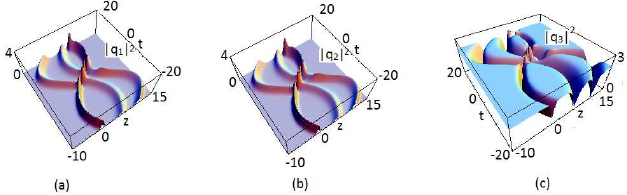}
\caption{Elastic interactions occurring at three points between bright 3-solitons in two components, 
 $\vert q_1\vert$ and $\vert q_2\vert $, 
and  dark 3-solitons in component, $\vert q_3\vert  $ ,   
with $k=\frac{\pi}{24}$, $\beta=-1$, $\sigma_1=\sigma_2=\sigma_3=1$
$ \eta_1=1. +.5 {\it i}$, $\eta_2=2.01 -.5 {\it i}$, $\eta_3=1.1 +{\it i}$, $\chi_3=1+{\it i}$ , $\xi_3=1$ , 
$\alpha_{11}= \alpha_{12}=\alpha_{13}=\alpha_{21}=\alpha_{22}=\alpha_{23}=1$
$\sigma(t)=.001$, $\Gamma(t)=.0005.$.}
\label{fig:elast}       
\end{figure}

\begin{figure}[htbp]
\centerline{\includegraphics[width=1.0\columnwidth]{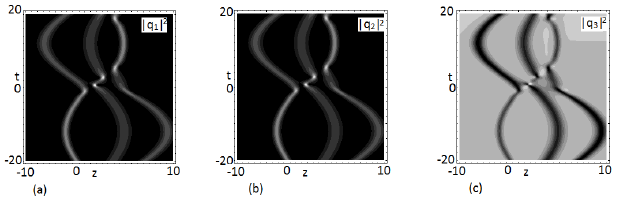}}
\caption{Contour plot of elastic interactions occurring at three points  between  bright 3-solitons in two components,  $\vert q_1\vert$ and $\vert q_2\vert $, 
and  dark 3-solitons in component, $\vert q_3\vert  $ ,   
with same set of parameters as in Fig \ref{fig:elast}}
\label{fig:elastc}
\end{figure}

\br
\label{EL2c}
\begin{aligned}
\alpha_1^{(2)}(t) & = \alpha_2^{(1)}(t) = \alpha_3^{(1)}(t)=0; \\
\alpha_1^{(1)}(t) &\ne 0; \alpha_2^{(2)}(t)\ne 0;   \alpha_3^{(2)}(t) \ne 0
\end{aligned}
\er
the amplitude of each soliton in each component remain same before and after interaction, That is
 \be
\label{ELint}
{\cal A}_j^{1-}= {\cal A}_j^{1+}; \quad   {\cal A}_j^{2-}= {\cal A}_j^{2+}; \quad   {\cal A}_j^{3-}= {\cal A}_j^{3+}. 
\ee
This is the case of an  elastic interaction \cite{SN15}. 
However, each solitons in each component suffer a 
phase-shift due to interaction, that is  $\Phi_j^{l-} - \Phi_j^{l+} \ne 0$ (for $l=1,2,3$). 
We use MATHEMATICA to plot the results of our analysis. To plot the figures we have chosen $D(t)= e^{\sigma t}cos(kt)$
and R(t)=cos(k t); consequently the gain coefficient $\Gamma(t) $ is nonvanishing.
Fig.\ref{fig:elast}(a, b) and Fig.\ref{fig:elastc}(a, b) show an example of periodic elastic interactions 
occurring at three points between three bright solitons  moving with different 
velocities, in  components $|q_1|$, $|q_2|$ and  
between three dark solitons in  component  $|q_3|$  Fig.(\ref{fig:elast}, \ref{fig:elastc}) (c), 
with $\alpha_1^{(1)}(t)$, $\alpha_1^{(2)}(t)$, $\alpha_2^{(1)}(t)$, 
$\alpha_2^{(2)}(t)$,  $\alpha_3^{(1)}(t)$ and 
$\alpha_3^{(2)}(t)$,  satisfying the condition \ref{EL1}. The trajectories of  solitons  resemble to three
synchronized damped simple harmonic oscillators, oscillating with their velocity changing  periodically between maxima (the point 
where the two soliton interact) and  minima.  
The soliton amplitude of each  soliton remains same after each interaction  but  changes in 
phases occur. However, after two successive  cycle of interactions net phase-shift of each soliton turns out to be zero. 
The trajectories of solitons in both the components remain identical.

On the other hand  if none of the eqs. \ref{EL1} - \ref{EL2c}  is  satisfied then, soliton amplitude changes after 
interaction. That is, 
\be
\label{ELint1}
{\cal A}_j^{1-} \ne {\cal A}_j^{1+}; \quad {\cal A}_j^{2-} \ne {\cal A}_j^{2+};
\quad {\cal A}_j^{3-} \ne {\cal A}_j^{3+}  
\ee
which is an inelastic interaction \cite{SN15}. However, the asymptotic analysis of dark solitons eqs. 
\ref{ASD1}, \ref{ASD2} show that the amplitudes of solitons do not change after interaction. That is, 
dark soliton interactions in multi-component NLS model  is similar to  scalar NLS model.

\begin{figure}[htbp]
\centerline{\includegraphics[width=1.0\columnwidth]{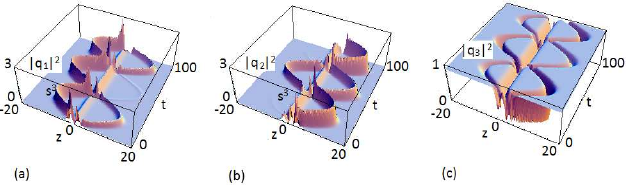}}
\caption{Inelastic Soliton interactions  between  bright 3-solitons in two components,  
$\vert q_1\vert$ and $\vert q_2\vert $, 
satisfying the conditions of a " NOT" like logic gate repeatedly
and  dark 3-solitons in component, $\vert q_3\vert  $ ,   
with 
$ \eta_1=1.2 +{\it i}$, $\eta_2=1.01$, $\eta_3=-1 - {\it i}$, 
 $\alpha_{11}= \alpha_{12}=\alpha_{23}=1$, $\alpha_{21}=\alpha_{22}=2$, $\alpha_{13}=8.$}
\label{fig:not1}
\end{figure}

\begin{figure}[htbp]
\centerline{\includegraphics[width=1.0\columnwidth]{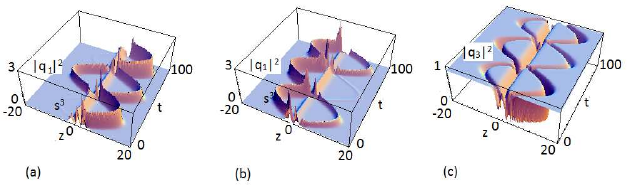}}
\caption{ Inelastic Soliton interactions  between  bright 3-solitons in two components,  
$\vert q_1\vert$ and $\vert q_2\vert $, 
 satisfying the conditions of a " NOT" logic gate repeatedly 
and  dark 3-solitons in component, $\vert q_3\vert  $ , 
with  
$ \eta_1=1.2 +{\it i}$, $\eta_2=1.01$, $\eta_3=-1 - {\it i}$,
 $\alpha_{11}= \alpha_{12}=2$, 
$\alpha_{21}=\alpha_{22}=\alpha_{13}=1$, $\alpha_{23}=8$.}
\label{fig:not2}
\end{figure}

Fig.(\ref{fig:not1}, \ref{fig:not2}) are two examples of inelastic 
interactions between three bright solitons in  components $|q_1|$, $|q_2|$ between three 
dark solitons in  components  $|q_3|$,  moving with arbitrarily chosen relative velocities,  and 
$\alpha_1^{(1)}(t)$,  $\alpha_1^{(2)}(t)$, $\alpha_2^{(1)}(t)$, $\alpha_2^{(2)}(t)$, 
$\alpha_3^{(1)}(t)$, $\alpha_3^{(2)}(t)$ are in accordance with the condition eq.\ref{ELint1}. 
In component $|q_1|$,  one of the  solitons
gains energy after interaction and in component $|q_2|$ it looses energy after interaction.   
The energy sharing  and phase-shift are however, temporary.  In a cycle of interactions net inter-component  energy transfer and  
soliton phase shift become zero, which is different from the inelastic interaction in a Manakov two soliton interaction
\cite{{Radhakrishnan97},{KANNA2016}}.  
This is interesting and we may describe this as if an input signal is  passing through two successive 'NOT' gates and ultimately 
remained unchanged. 

\begin{figure}[htbp]
\centerline{\includegraphics[width=1.0\columnwidth]{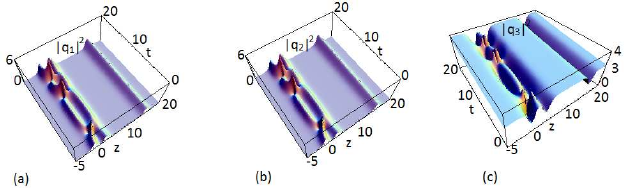}}
\caption{ Plot of bound bright 3-solitons in  two components, 
 $\vert q_1\vert$ and $\vert q_2\vert $ and  dark 3-solitons in component, $\vert q_3\vert  $ ,  
with $k= \frac{\pi}{36}$, $\gamma=-1 $, $\sigma_1 =\sigma_2=\sigma_3=1$, 
$ \eta_1=1$, $\eta_2=1.01$, $\eta_3=2$,
$\chi_3=1+{\it i}$ , $\xi_3=1$ , $\alpha_{11}= \alpha_{12}=\alpha_{23}=\alpha_{21}
=\alpha_{22}=\alpha_{13}=1$, $\sigma =0.1$ ,$\Gamma(t)=0.05$.}
\label{fig:3SSa}
\end{figure}

\begin{figure}[htbp]
\centerline{\includegraphics[width=1.0\columnwidth]{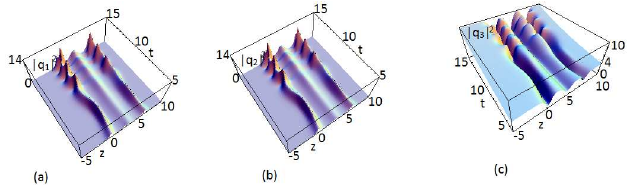}}
\caption{ Plot of bound bright 3-solitons in  two components, 
 $\vert q_1\vert$ and $\vert q_2\vert $ and  dark 3-solitons in component, $\vert q_3\vert  $ ,  
with $k= \frac{\pi}{36}$, $\gamma=-1 $, $\sigma_1 =\sigma_2=\sigma_3=1$, 
$ \eta_1=1$, $\eta_2=3$, $\eta_3=3.1$,
$\chi_3=1+{\it i}$ , $\xi_3=1$ , $\alpha_{11}= \alpha_{12}=\alpha_{23}=\alpha_{21}
=\alpha_{22}=\alpha_{13}=1$, $\sigma =0.1$ ,$\Gamma(t)=0.05$.}
\label{fig:3SSb}
\end{figure}
  
\begin{figure}[htbp]
\centerline{\includegraphics[width=1.0\columnwidth]{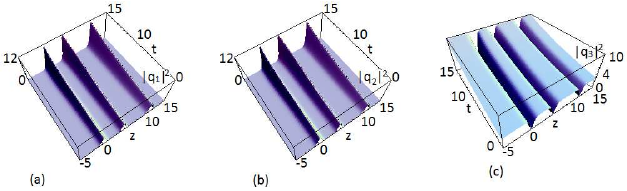}}
\caption{ Plot of bound bright 3-solitons in  two components, 
 $\vert q_1\vert$ and $\vert q_2\vert $ and  dark 3-solitons in component, $\vert q_3\vert  $ ,  
with $k= \frac{\pi}{36}$, $\gamma=-1 $, $\sigma_1 =\sigma_2=\sigma_3=1$, 
$ \eta_1=2$, $\eta_2=2.1$, $\eta_3=2.01$,
$\chi_3=1+{\it i}$ ,  $\xi_3=1$ ,  $\alpha_{11}= \alpha_{12}=\alpha_{23}=\alpha_{21}
=\alpha_{22}=\alpha_{13}=1$,  $\sigma =0.1$ , $\Gamma(t)=0.05$.}
\label{fig:3SSc}
\end{figure}
\begin{figure}[htbp]
\centerline{\includegraphics[width=1.0\columnwidth]{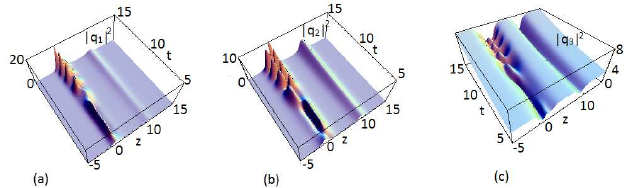}}
\caption{Plot of energy sharing  bound bright 3-solitons in  two components, 
 $\vert q_1\vert$ and $\vert q_2\vert $ and  dark 3-solitons in component, $\vert q_3\vert  $ ,  
with $k= \frac{\pi}{36}$, $\gamma=-1 $, $\sigma_1 =\sigma_2=\sigma_3=1$, 
$ \eta_1=1$, $\eta_2=1.01$, $\eta_3=2.01$,
$\chi_3=1+{\it i}$ , $\xi_3=1$ , $ -\alpha_{11}= \alpha_{12}=\alpha_{23}=\alpha_{21}
=\alpha_{22}=\alpha_{13}=1$, $\sigma =0.1$ ,$\Gamma(t)=0.05$.}
\label{fig:3SSBa}
\end{figure}
\begin{figure}[htbp]
\centerline{\includegraphics[width=1.0\columnwidth]{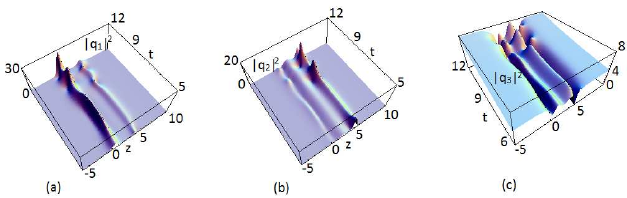}}
\caption{ Plot of energy sharing bound bright 3-solitons in  two components, 
 $\vert q_1\vert$ and $\vert q_2\vert $ and  dark 3-solitons in component, $\vert q_3\vert  $ ,  
with $k= \frac{\pi}{36}$, $\gamma=-1 $, $\sigma_1 =\sigma_2=\sigma_3=1$, 
$ \eta_1=1$, $\eta_2=2.2$, $\eta_3=2.21$,
$\chi_3=1+{\it i}$ , $\xi_3=1$ , $\alpha_{11}= \alpha_{12}=-\alpha_{23}=\alpha_{21}
=\alpha_{22}=\alpha_{13}=1$, $\sigma =0.1$ ,$\Gamma(t)=0.05$.}
\label{fig:3SSBb}
\end{figure}
  
\begin{figure}[htbp]
\centerline{\includegraphics[width=1.0\columnwidth]{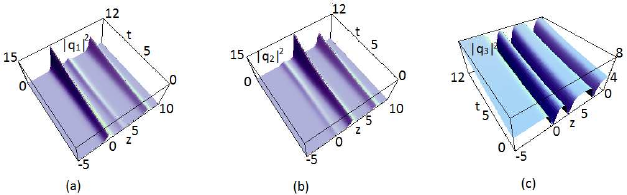}}
\caption{ Plot of energy sharing bound bright 3-solitons in  two components, 
 $\vert q_1\vert$ and $\vert q_2\vert $ and  dark 3-solitons in component, $\vert q_3\vert  $ ,  
with $k= \frac{\pi}{36}$, $\gamma=-1 $, $\sigma_1 =\sigma_2=\sigma_3=1$, 
$ \eta_1=2$, $\eta_2=2.1$, $\eta_3=2.01$,
$\chi_3=1+{\it i}$ ,  $\xi_3=1$ ,  $\alpha_{11}= \alpha_{12}=-\alpha_{23}=\alpha_{21}
=\alpha_{22}=\alpha_{13}=1$,  $\sigma =0.1$ , $\Gamma(t)=0.05$.}
\label{fig:3SSBc}
\end{figure}

\begin{figure}[htbp]
\centerline{\includegraphics[width=1.0\columnwidth]{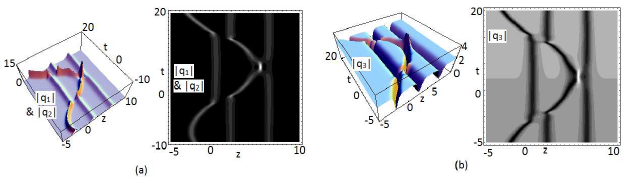}}
\caption{ Plot of  Soliton interactions  between a bright 2-soliton complex, with an accelerating soliton,
 in   components,  $| q_1|$ and $| q_2| $.(bright soliton),   
with $ k= \frac{\pi}{36}$, $\gamma= -1 $, $\sigma_1 =\sigma_2=\sigma_3=1$, 
$\eta_1=2$, $\eta_2= 2.1 $, $\eta_3=3.01+ {\it i}$,  $\chi_3=1 +{\it i}$, $\xi_3=1$,
 $\alpha_{11}= \alpha_{12} =\alpha_{21}=\alpha_{22}=\alpha_{13}=\alpha_{23}=1$,
$\sigma =0.01 $ , $\Gamma(t)=0.005$.}
\label{fig:3ss2bndbr}
\end{figure}

\begin{figure}[htbp]
\centerline{\includegraphics[width=1.0\columnwidth]{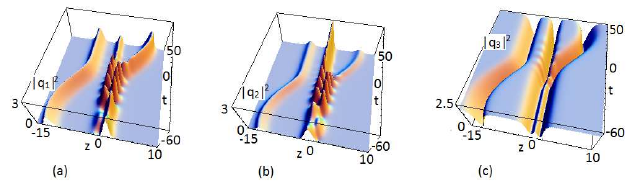}}
\caption{  Plot of  Soliton interactions  between a   2-soliton complex, which changes to a breather type soliton 
after each interaction and an accelerating soliton, in  components,  $| q_1|$, $| q_2| $ (bright solitons) and     $ | q_3 |  $ 
(dark soliton),   
with $ k= \frac{\pi}{96}$, $\gamma= -1 $, $\sigma_1 =\sigma_2=\sigma_3=1$, 
$ \eta_1=-1$, $\eta_2= -1.6$, $\eta_3=- 1 +.2 {\it i}$,  $\chi_3=1 +{\it i}$, $\xi_3=1$,
$ \delta_2= \delta_4 = 0$,  that is,
 $\alpha_{11}= \alpha_{12} =-\alpha_{21}=\alpha_{22}=1$,  $\alpha_{13}=2$, $\alpha_{23}=1$,
$\sigma =0.006 $ , $\Gamma(t)=0.003$.}
\label{fig:3SBREATHER}
\end{figure}

\begin{figure}[htbp]
\centerline{\includegraphics[width=1.0\columnwidth]{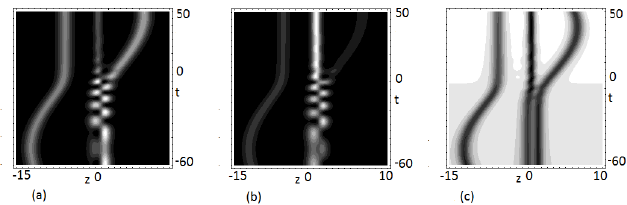}}
\caption{Contour Plot of Soliton interactions  between a  2-soliton complex,  with an accelerating soliton, 
in  components,  $| q_1|$, $ | q_2| $ (bright solitons) and     $ | q_3 |  $ (dark soliton) ,   
with same set of parameters as in Fig.\ref{fig:3SBREATHER}}
\label{fig:3SBREATHERC}
\end{figure}

\begin{figure}[htbp]
\centerline{\includegraphics[width=1.0\columnwidth]{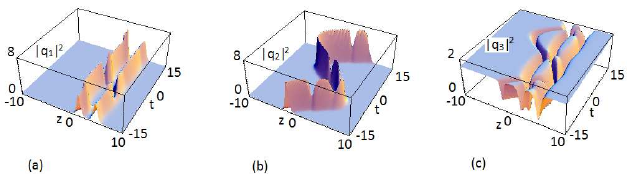}}
\caption{ Plot of Soliton interactions  between   2-soliton complex  and an accelerating soliton, 
in  components, $ | q_1|$, $|q_2| $ (bright solitons) and   $ |q_3|$ (dark soliton),   
with $k= \frac{\pi}{12}$, $\gamma= 1 $, $\sigma_1 =\sigma_2=\sigma_3=1$, 
$ \eta_1=2.1$, $\eta_2=2$, $\eta_3=2.5 -{\it i}$,  $\chi_3=1 -{\it i}$, $\xi_3=1$,
$ \delta_3= \delta_7=\delta_6=\delta_8=0$,  that is,
 $\alpha_{11}= \alpha_{12}=1$, $\alpha_{21}=\alpha_{22}=0$,  $\alpha_{13}=0$, $\alpha_{23}=1$,
 $\sigma =0.006$ ,$\Gamma(t)=0.003$.}
\label{fig:bound-delta}
\end{figure}

\begin{figure}[htbp]
\centerline{\includegraphics[width=1.0\columnwidth]{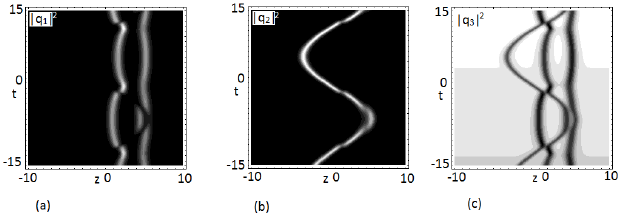}}
\caption{ Contour Plot of Soliton interactions  between  a 2-soliton complex  and an accelerating soliton, 
in   components, $ | q_1 |$, $ | q_2| $ (bright solitons) and   $ | q_3|  $ (dark soliton),   
with same set of parameters as in Fig.\ref{fig:bound-delta}}
\label{fig:bound-deltac}
\end{figure}

\begin{figure}[htbp]
\centerline{\includegraphics[width=1.0\columnwidth]{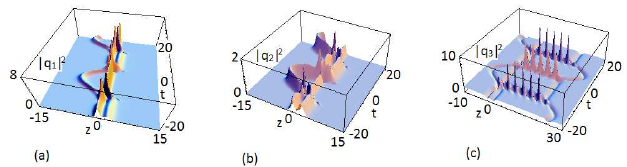}}
\caption{ Soliton interactions  between  a bright 2-soliton complex  with an accelerating soliton in
$\vert q_1\vert$, $\vert q_2\vert $ and two dark solitons complex  with an accelerating soliton in $\vert q_3\vert  $ ,   
with $k= \frac{\pi}{12}$, $\gamma=-1 $, $\sigma_1 =\sigma_2=\sigma_3=1$, 
$ \eta_1=1 + 2{\it i}$, $\eta_2=2.01$, $\eta_3=1 	$, $\chi_3=1 -{\it i}$, $\xi_3=1$,
 $\alpha_{11}= \alpha_{22}=\alpha_{23}=1$, $\alpha_{21}=\alpha_{13}=\alpha_{12}=-1$,
 $\sigma =0.005$ ,$\Gamma(t)=0.0025$.}
\label{fig:longph}
\end{figure}

\begin{figure}[htbp]
\centerline{\includegraphics[width=1.0\columnwidth]{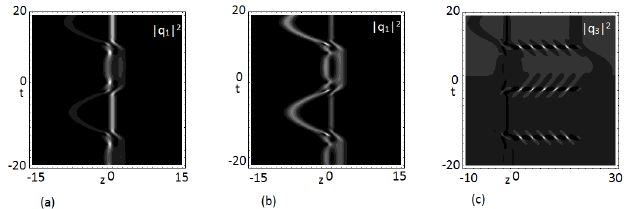}}
\caption{Contour Plot of Soliton interactions  between  a 2-soliton complex  with an accelerating soliton, effecting 
a large phase shift in all  components, $\vert q_1\vert$, $\vert q_2\vert $ (bright solitons) and 
 $\vert q_3\vert  $ ( dark soliton),   
with same set of parameters as in Fig.\ref{fig:longph}}
\label{fig:longphc}
\end{figure} 

\section{THREE SOLITON BOUND STATE}\label{complex}

In three soliton bound state we consider  two cases. In case one  solitons have no relative velocities among themselves and 
in case two, a two solitons complex interact with a third soliton moving with a relative velocity.
We analyze the bound soliton interactions using  asymptotic analysis, wherever applicable as well 
as graphical analysis. When all three solitons are comoving, that is the solitons 
do not move sufficiently
apart  from each other  asymptotically and hence an effective asymptotic analysis is difficult.
We identify  a correlations, among soliton parameters, which  distinguishes different types of interactions among 
the solitons. First we consider 
the three comoving solitons, with  components
satisfying one of the following set of conditions.   \\
Case {\bf  A},
\be
\alpha_{11}=\alpha_{21}=\alpha_{12}= \alpha_{22}=\alpha_{13}=\alpha_{23}= a
\label{CASE1}
\ee
First suffix  on $\alpha$  denotes component  and second  suffix  denotes soliton number(see Appendix A). "a" is a real constant.
The condition  eq.\ref{CASE1} comply with the condition for elastic collision eq.\ref{EL1}, given in the previous section.   
Fig.\ref{fig:3SSa} (a,b) shows an instance when the solitons  are satisfying \ref{CASE1} as well as the following 
condition, $Re(\eta_3) > Re(\eta_1),  Re(\eta_2)$ (interchangeable), where  $|\eta_1|$ and $ |\eta_2|$ are comparable.
The figure shows the trajectories of three bright solitons  
in an  induced medium (induced by the combined power of three solitons).  Two of the solitons 
of comparable  intensities (solitons with  $\eta_1, \eta_2$) interact periodically. 
The  remaining soliton, with higher intensity,  however, remains isolated  and  sticks to a straight trajectory until the gain 
becomes very high. Dark solitons  Fig.\ref{fig:3SSa}(c) maintain nearly a similar profile as that of  bright solitons. 

Fig.\ref{fig:3SSb}(a,b)  shows another instance of  bound  soliton interaction, where the solitons satisfy eq.\ref{CASE1} 
as well as follow the condition, $Re(\eta_1) < Re(\eta_2),  Re(\eta_3)$,  where  $\eta_2;  \eta_3$ are comparable.
In this case two of the  solitons of stronger  
intensities (soliton with $\eta_2,  \eta_3$) are seen to interact periodically in this case. 
The interactions become stronger with gain. 
The trajectory of the soliton with lower intensity, however  remains straight. The dark solitons  
Fig.\ref{fig:3SSb}(c), once again maintain nearly a similar profile as that of  the bright solitons. 
Fig.\ref{fig:3SSc} displays the third 
example of  bound soliton interactions of three solitons  when the soliton components satisfy eq.\ref{CASE1} and 
have nearly same intensities. The bright as well as the dark 
solitons in this case show no sign of interactions even though the intensity increases due to gain 
and they maintain a similar  straight trajectories.

The above analysis shows that in all the three examples  bright soliton interactions and and their trajectories  
in first component are identical to that in second component. That is, just like scalar solitons the energy of each soliton 
is conserved in each component. 
    
Case {\bf  B.}
\be
\begin{aligned} 
\alpha_{21} &=\alpha_{12}= \alpha_{22} =\alpha_{13}
=\alpha_{23}= a \\
\alpha_{11} &= -a; \quad  or\\
 \alpha_{11} &=\alpha_{12}= \alpha_{22} =\alpha_{13}
=\alpha_{23}= a \\
\alpha_{21}&=-a; \quad \\
\end{aligned}
\label{BINEL}
\ee
These conditions  also imply that the following parameters become zero, thus making the three soliton expression simpler
( see Appendix A),
\br
\begin{aligned}
\delta_2  &= \delta_2^* = \delta_3=\delta_3^*=0 \\
\gamma_2  &= \gamma_2^* =\gamma_3=\gamma_3^*=0 \\
\varrho_3 &= \varrho_3^* =\varrho_6 =\varrho_6^*=0\\
\rho_3  &=  \rho_3^* = \rho_6 =\rho_6^* ; \quad  \beta_{j7}=0\\
\end{aligned}
\er
Notice that, eq. \ref{BINEL}  does not match with any one of the conditions eqs.\ref{EL1} -\ref{EL2c}
for elastic interactions. A similar restriction  are imposed on $\eta_1,\eta_2$ and $\eta_3$ as earlier but  now along
with eq. \ref{BINEL}.  First we consider the case,  $Re(\eta_3) > Re(\eta_1),  Re(\eta_2)$ (interchangeable), 
where $|\eta_1|$ and  $|\eta_2|$ are comparable.
The bound bright solitons,  in Fig.\ref{fig:3SSBa}(a) 
are not only interacting,  but forming a  breather like localized structure, alternately shifting 
position between the trajectories  of each soliton.  Third soliton  maintain a straight trajectory 
until the gain  ($e^{\sigma t}$) becomes too high, then all three 
solitons are subjected to attraction and repulsion.
Fig.\ref{fig:3SSBa}(b) showing the soliton trajectories in the 
second component but in this case the trajectories rather than being identical to the first component are 
complementary in nature. This suggests that there is an energy sharing process involved in this interaction.
The bound dark solitons in Fig.\ref{fig:3SSBa}(c)  are also  interacting but no energy
exchange process is involved. The remaining (non interacting) dark  soliton  maintains a straight trajectory. 
Similarly when the  condition eq.\ref{BINEL} along with the condition,
 $Re(\eta_1) < Re(\eta_2),  Re(\eta_3)$, where  $|\eta_2|$ and $|\eta_3|$ are comparable are imposed on the solitons 
another instance of energy sharing between the components is noticed in Fig.\ref{fig:3SSBb}(a,b). The 
trajectories of the interacting solitons are similar to that in Fig.\ref{fig:3SSb}(a,b), but the intercomponent 
energy sharing is more distinct. 
When all the solitons are of comparable intensities and they follow \ref{BINEL} we notice an interesting 
phenomena. 
Fig. \ref{fig:3SSBc}(a, b) shows that despite the fact, that the soliton trajectories remain parallel and there is no sign of 
interaction, but, the energy exchange between the components  still occur. 
The solitons in second component once again happen to be complementary to the solitons in the first component.
Dark solitons in Fig.\ref{fig:3SSBc}(c), on the other  hand follow  straight  trajectories similar to that in Fig.\ref{fig:3SSc}.\\

Similar to Case {\bf B} two more cases may  be obtained when 
($\alpha_{11}=\alpha_{21}=\alpha_{12}= \alpha_{22}=-\alpha_{13}=\alpha_{23}= a$), 
 or  
($\alpha_{11}=\alpha_{21}=\alpha_{12}= \alpha_{22}=\alpha_{13}=-\alpha_{23}= a$).  
Then the following parameters become zero,
\br
\begin{aligned}
\delta_3 &= \delta_3^* = \delta_6=\delta_6^*=0 \\
\gamma_3  &= \gamma_3^* =\gamma_6=\gamma_6^*=0 \\
\varrho_2 &= \varrho_4^* =\varrho_3 =\varrho_7^*=0\\
\rho_2  &= \rho_4^* = \rho_3 =\rho_3^*; \quad \beta_{j3}=0
\end{aligned}
\er

and 
($\alpha_{11}=\alpha_{21}=-\alpha_{12}= \alpha_{22}=\alpha_{13}=\alpha_{23}= a$),  
or  
($\alpha_{11}=\alpha_{21}=\alpha_{12}= -\alpha_{22}=\alpha_{13}=\alpha_{23}= a$)  
Then the following parameters become zero,
\br
\begin{aligned}
\delta_2 &= \delta_2^* = \delta_6=\delta_6^*=0 \\
\gamma_2  &= \gamma_2^* =\gamma_6=\gamma_6^*=0 \\
\varrho_2 &= \varrho_4^* =\varrho_6 =\varrho_8^*=0\\
\rho_2  &= \rho_4^* = \rho_6 =\rho_8^*; \quad \beta_{j5}=0
\end{aligned}
\er
Then also we expect a similar trajectories of solitons as described earlier (Case B). 
Notice that the  interactions among the solitons  in bound state is  medium induced interactions and 
not a collision arising out of relative velocity of the solitons. However, the inter component 
energy exchange, a phenomena normally mentioned in the literature  for  colliding solitons moving with a relative velocity
is prevalent also in case of bound solitons. \\

Interaction of a two soliton complex and an accelerating soliton, on the other hand is more eventful as will be explained 
using the asymptotic analysis. Here, asymptotic analysis is applied to three soliton solutions, where a   bright two soliton complex, 
$S^{12}$ (say), is interacting  with another soliton $S^3$(say), which has  a relative velocity with respect to $S^{12}$.  
 $S^{12}$ and  $S^3$ remain  separated asymptotically.

Thus  as $z \rightarrow - \infty$,  asymptotic expressions for $S^{12}$ and $S^{3}$ are,
\br
\begin{aligned}
S^{12} &= \frac{\alpha^{(j)}_1 (t)e^{\theta_1} +\alpha^{(j)}_2(t) e^{\theta_2} + 
\beta_{j1} e^{\theta_1 +\theta_1^*+\theta_2 } +\beta_{j2} e^{\theta_1 +\theta_2^*+\theta_2 }}
{1+  \delta_1  e^{\theta_1 +\theta_1^*}  + \delta_2  e^{\theta_1 +\theta_2^*} +\delta_2^*  e^{\theta_2 +\theta_1^*}
+ \delta_5  e^{\theta_2 +\theta_2^*}  + \Upsilon  e^{\theta_1 +\theta_1^* +\theta_2 +\theta_2^*} };\\ 
&\theta_3 +\theta_3^* \rightarrow -\infty \\
S^3 &= \frac{ \varsigma_{1}^{(j)} }{ \sqrt{ \varrho_1  \Upsilon }} Sech(\frac{ \theta_3 +\theta_3^*  
+ ln (\frac{\Upsilon}{\varrho_1 } )}{2}); \quad \quad \theta_1 +\theta_1^* +\theta_2 +\theta_2^*  
\rightarrow +\infty 
\end{aligned}
\er
and as $z \rightarrow + \infty$, the expressions are 
\br
\begin{aligned}
S^{12}_j &= \frac{\beta_{j6} e^{\theta_1} + \beta_{j9}  e^{\theta_2} + 
\varsigma_{2}^{(j)} e^{\theta_1 +\theta_1^*+\theta_2 } +\varsigma_{3}^{(j)} e^{\theta_1 +\theta_2^*+\theta_2 } }
{\delta_9 +  \varrho_5  e^{\theta_1 +\theta_1^*}  + \varrho_6  e^{\theta_1 +\theta_2^*} 
+ \varrho_6^*  e^{\theta_2 +\theta_1^*}  + \varrho_9 e^{\theta_2 +\theta_2^*}  +
 \Upsilon  e^{\theta_1 +\theta_1^* +\theta_2 +\theta_2^*} }; \\
&\theta_3 +\theta_3^* \rightarrow + \infty \\
S^{3}_j &= \frac{ \alpha^{(j)}_3(t) }{ \sqrt{ \delta_9 }} Sech(\frac{ \theta_3 +\theta_3^*  
+ ln (\delta_9)}{2}); \qquad
\theta_1 +\theta_1^* + \theta_2 +\theta_2^*  \rightarrow - \infty 
\end{aligned}
\er

Similarly  consider the interaction of a  dark two  soliton complex, $X^{12}$ (say) and  a third dark soliton  $X^3$, which is
moving at a relative speed with respect to  $X^{12}$. 

Thus asymptotic expressions for the dark solitons are, as $z \rightarrow - \infty$,
\br
\begin{aligned}
X^{12} & = \frac{ 1+ \gamma_1^{(3)} e^ {\theta_1 + \theta_1^*} 
+  \gamma_2^{(3)} e^{\theta_1 + \theta_2^*} + 
\gamma_4^{(3)} e^{\theta_2 + \theta_1^*} +  e^{\theta_2 + \theta_2^*}(\gamma_5^{(3)}  +
\nu^{(3)} e^{\theta_1  + \theta_1^* })}{
( 1 + \delta_1 e^ {\theta_1 + \theta_1^*} +  \delta_2 e^{\theta_1 + \theta_2^*} + 
\delta_2^* e^{\theta_2 + \theta_1^*} + \delta_5 e^{\theta_2 + \theta_2^*} +
\Upsilon e^{\theta_1 +  \theta_2 + \theta_1^*+ \theta_2^*} )}\\
&\times\chi^{(3)}(t) e^{\phi^{(3)} };\qquad
\theta_3 + \theta_3^*  \rightarrow - \infty  \\
X^{3} & =  \chi^{(3)}(t) e^{\phi^{(3)} + \frac{\theta_3 - \theta_3^*}{2}}
 \frac{\rho_1^{(3)}}{\varrho_1}
( 1 + \frac{\gamma_9^{(3)}}{\delta_9} + (\frac{\gamma_9^{(3)}}{\delta_9} -1)  
Tanh( \frac{\theta_3 + \theta_3^* +ln({\frac{\Upsilon}\varrho_1})}{2}); \\ 
\theta_1 &+ \theta_1^*  + \theta_2 + \theta_2^*  \rightarrow + \infty 
\end{aligned}
\er
and as $z \rightarrow  + \infty $,  
\br
\begin{aligned}
X^{12} & =  \frac{\gamma_9^{(3)} + \rho^{(3)}_5 e^ {\theta_1 + \theta_1^*} 
+  \rho_6^{(3)} e^{\theta_1 + \theta_2^*} + \rho_8^{(3)} e^{\theta_2 + \theta_1^*} 
+ e^{\theta_2 + \theta_2^*} (\rho_9^{(3)}  +
\nu^{(3)} e^{\theta_1 + \theta_1^*} )}{
( \delta_9 + \varrho_5 e^ {\theta_1 + \theta_1^*} +  \varrho_6 e^{\theta_1 + \theta_2^*} + 
\varrho_6^* e^{\theta_2 + \theta_1^*} + \varrho_9 e^{\theta_2 + \theta_2^*} +
\Upsilon e^{\theta_1 +  \theta_2 + \theta_1^*+ \theta_2^*} )}\\
&\times \chi^{(3)}(t) e^{\phi^{(3)} }; \qquad
\theta_3  + \theta_3^*  \rightarrow +  \infty \\
X^{3} & =  \chi^{(3)}(t) e^{\phi^{(3)} + \frac{\theta_3 - \theta_3^*}{2}}
( 1 + \frac{\gamma_9^{(3)}}{\delta_9} + (\frac{\gamma_9^{(3)}}{\delta_9} -1)  
Tanh( \frac{\theta_3 + \theta_3^* +ln(\delta_9)}{2});\\
 &~ \theta_1 + \theta_1^*  + \theta_2 + \theta_2^*  \rightarrow - \infty 
\end{aligned}
\er

Once again consider, the solitons are such that, the components are satisfying 
eq. \ref{CASE1}.  The values of  $\eta_1$, $\eta_2$ and $\eta_3$ are chosen such that two of them form a
comoving bright soliton complex $S^{12}$ (dark soliton complex, $X^{12}$ ) and interact with  
the other bright soliton $S^3$ (dark soliton $X^3$).  
Fig. \ref{fig:3ss2bndbr} (a)  demonstrate  a case where two bright solitons with 
parallel trajectories interact with an accelerating soliton. First interaction causes a shift in trajectories  
(phase shift) of both the interacting soliton.  The moving soliton then decelerate and interact with the second 
soliton just when its velocity reached minimum. The point is marked with a bright spot. The maxima of two 
solitons just get merged at that point. Both the components $|q_1|$ and  $|q_2|$  show identical profile of solitons.  
Fig. \ref{fig:3ss2bndbr}(b)  shows the dark soliton interactions under the 
same conditions as stated in eq. \ref{CASE1} The first interaction causes a normal phase shift. 
Interestingly  at the second interaction point  the merging of two solitons is marked with an increase in intensity, 
which exceeds the background intensity. \\

Fig. \ref{fig:3SBREATHER} and\ref{fig:3SBREATHERC} demonstrate another instance of bound soliton interaction, 
where  two bright comoving soliton complex $S^{12}$ (dark soliton complex, $X^{12}$ ) 
interact with  $S^3$ ($X^3$), having a relative velocity with respect to $S^{12}$ ($X^{12}$) .   
The soliton components satisfy,
($\alpha_{11}=-\alpha_{21}= \alpha_{12}=\alpha_{22}= -\frac{1}{2}\alpha_{13}=\alpha_{23}=1$), 
that is, they do not satisfy any one of the  conditions  eqs.\ref{EL1}- \ref{EL2c}.
Fig. \ref{fig:3SBREATHER} (a, b),  \ref{fig:3SBREATHERC}(a, b) show that,  interaction  brings the two 
comoving soliton complex  nearer effecting a large phase shift. They form a breather like structure as a 
result of the mutual interactions.  
In the subsequent interaction the   solitons are  again separated sufficiently so that
mutual interaction between them disappear and this 
phenomena continue in a periodic manner. The interaction of dark solitons causes a similar effect on their phase(position), moving 
the bound solitons closer and  farther at regular intervals, Fig. \ref{fig:3SBREATHER} (c), \ref{fig:3SBREATHERC}(c). 
 
Fig.(\ref{fig:bound-delta}, \ref{fig:bound-deltac}) (a, b, c)  shows a special case, where the soliton components 
($\alpha_{11}=\alpha_{12}=\alpha_{23}=1;   \alpha_{21}=\alpha_{22}= \alpha_{13}=0 $ )
satisfy the conditions of an elastic collision, namely eq.\ref{EL2bc}. The values of $\eta$ are chosen as shown in the figure.
In this interactions soliton  ($S^3$) completely disappears from component $|q_1|$ and  the soliton complex $S^{12}$ disappears 
from component  $|q_2|$.  However, traces of interactions are noticed  in  their phase shifts.
Dark soliton on the other hand  exhibit a regular interaction of an accelerating 
soliton with mutually interacting bound solitons.

Finally we bring  an interesting phenomena where the soliton interaction causes a very large phase shift. The component of the solitons
are satisfying eq. \ref{EL1}, that is the interaction is an elastic one. 
Fig.\ref{fig:longph} \ref{fig:longphc}(c) show that 
during  the phase change dark solitons  shows a dramatic variation of intensities,  maintaining a unique  envelope shape. 
This produces equally spaced  fringes of bright and dark lines. The values of $\eta$' are as shown in the figure. 
Bright soliton Fig.\ref{fig:longph} (a, b),  on the other hand demonstrate that the interaction pushed 
solitons out of the frame, they reappear only during 
next cycle of interaction. 

\section{Conclusion}\label{Conclusion}

Three soliton solution of  a Lax integrable coupled NLS equation with variable coefficients is obtained using Hirota's bilinear 
method. 
Soliton complexes are studied using  
asymptotic as well as graphical analysis. 
It is shown  that two comoving solitons  of  sufficient 
amplitude are subjected to periodic attraction and repulsion.
The energy sharing phenomena 
which was earlier reported for colliding Manakov solitons has been noticed also in non 
colliding  comoving bound state solitons. 
A correlation between the bound state soliton interactions and the soliton parameters is shown through 
graphical analysis. One of the interesting phenomena 
noticed in three soliton bound state interaction  is the long phase shift of solitons. The same is explained on the 
basis of the asymptotic analysis.   Such observations have not been reported earlier either for two soliton or for three soliton. 
This allows us to draw  a conclusion that the  intercomponent energy exchange soliton collision is not a necessary condition.  
The analysis done in this paper are expected to 
be useful for the development of all optical logic gates, soliton communication systems as well as to understand
bound soliton complexes.

\begin{center}
{\bf Appendix }
\end{center}
Bright 3-soliton is in  components $|q_1|$ and $|q_2|$ and dark 3-soliton is in component $|q_3|$
\be
\label{2SS}
q_j = \frac{g_1^{(j)} + g_3^{(j)} + g_5^{(j)} }{1+ f_2 + f_4 +f_6}; \qquad (for \quad j=1,2)\nonu
\ee
\brs
\label{2SSP}
q_3 &=& \frac{g_0^{(3)}(1 + g_2^{(3)} +g_4^{(3)} + g_6^{(3)} )}{1+ f_2 +f_4 + f_6};  \\
\text{where,}&~& \\
{g_0}^{(3)} &=& \chi^{(3)}(t)  e^{\phi(3)}; \qquad 
{g_1}^{(j)} = \displaystyle{\sum_{k=1}^{3} \alpha_k^{(j)}(t) e ^{\theta_k} } \nonumber\\
{g_2}^{(3)} &=& \gamma_1^{(3)} e^{\theta_1 + {\theta_1}^*} + \gamma_2^{(3)} e^{\theta_1 + {\theta_2}^*} +  
\gamma_3^{(3)} e^{\theta_1 + {\theta_3}^*} \nonumber \\&~& +
 \gamma_4^{(3)} e^{\theta_2 + {\theta_1}^*}   
\gamma_5^{(3)} e^{\theta_2 + {\theta_2}^*} + \gamma_6^{(3)} e^{\theta_2 + {\theta_3}^*} \nonumber \\&~& +
\gamma_7^{(3)} e^{\theta_3 + {\theta_1}^*} + \gamma_8^{(3)} e^{\theta_3 + {\theta_2}^*} +
\gamma_9^{(3)} e^{\theta_3 + {\theta_3}^*} \nonumber 
\ers
\br
{g_3}^{(j)} &=& \beta_{j 1} e^{\theta_1 + \theta_2 + {\theta_1}^* } + \beta_{j 2} e^{\theta_1+\theta_2+{\theta_2}^*} +
\beta_{j 3} e^{\theta_1+\theta_2+{\theta_3}^*}\nonumber \\ 
&~& + \beta_{j 4} e^{\theta_1+\theta_3+{\theta_1}^*}  
\beta_{j 5} e^{\theta_1+\theta_3+{\theta_2}^*} + \beta_{j 6} e^{\theta_1+\theta_3+{\theta_3}^*} 
 \nonumber \\ & ~ & + 
\beta_{j 7} e^{\theta_2+\theta_3+{\theta_1}^*} + \beta_{j 8} e^{\theta_2+\theta_3+{\theta_2}^*}
\beta_{j 9} e^{\theta_2+\theta_3+{\theta_3}^*} \nonumber \\
{g_4}^{(3)} &=& \rho_1^{(3)} e^{\theta_1 + {\theta_2} +{\theta_1}^* + {\theta_2}^*}  +
 \rho_2^{(3)} e^{\theta_1 + \theta_2+ {\theta_1}^*  + {\theta_3}^*} \nonumber \\&~& +
\rho_3^{(3)} e^{\theta_1 + \theta_2 + {\theta_2}^* + {\theta_3}^*}  +
\rho_4^{(3)} e^{\theta_1 + \theta_3 + {\theta_1}^* + {\theta_2}^*} \nonumber \\&~&+
\rho_5^{(3)} e^{\theta_1 + \theta_3 + {\theta_1}^* + {\theta_3}^*} +
\rho_6^{(3)} e^{\theta_1 + \theta_3 + {\theta_2}^* + {\theta_3}^*} \nonumber \\ &~&+
\rho_7^{(3)} e^{\theta_2 + \theta_3 + {\theta_1}^* + {\theta_2}^*} +
\rho_8^{(3)} e^{\theta_2 + \theta_3 + {\theta_1}^* + {\theta_3}^*} \nonumber \\&~& +
\rho_9^{(3)} e^{\theta_2 + \theta_3 + {\theta_2}^* + {\theta_3}^*}\nonumber
\er
\br
{g_5}^{(j)} &=& \varsigma_{1}^{(j)}e^{\theta_1+\theta_2 + \theta_3 + {\theta_1}^* + {\theta_2}^* } + 
\varsigma_{2}^{(j)}e^{\theta_1+\theta_2 + \theta_3 + {\theta_1}^* + {\theta_3}^* }  \nonumber \\ &~&+
\varsigma_{3}^{(j)}e^{\theta_1+\theta_2 + \theta_3 + {\theta_2}^* + {\theta_3}^* } \nonumber \\
{g_6}^{(3)} &=& \nu^{(3)} e^{\theta_1 + {\theta_2} +
{\theta_3} +{\theta_1}^* + {\theta_2}^*+ {\theta_3}^*}\nonumber\\
f_2 &=& \delta_1 e^{\theta_1 + {\theta_1}^*} + \delta_2 e^{\theta_1 + {\theta_2}^*} +  
\delta_3 e^{\theta_1 + {\theta_3}^*} + \delta_2^* e^{\theta_2 + {\theta_1}^*} \nonumber \\&~& + 
\delta_5 e^{\theta_2 + {\theta_2}^*} + \delta_6 e^{\theta_2 + {\theta_3}^*} +
\delta_3^* e^{\theta_3 + {\theta_1}^*} + \delta_6^* e^{\theta_3 + {\theta_2}^*} \nonumber \\&~& +
\delta_9 e^{\theta_3 + {\theta_3}^*} \nonumber\\
f_4 &=& \varrho_1 e^{\theta_1 + {\theta_2} +{\theta_1}^* + {\theta_2}^*} +
\varrho_2 e^{\theta_1 + \theta_2+ {\theta_1}^*  + {\theta_3}^*} \nonumber \\&~& +
\varrho_3 e^{\theta_1 + \theta_2 + {\theta_2}^* + {\theta_3}^*} +
\varrho_2^* e^{\theta_1 + \theta_3 + {\theta_1}^* + {\theta_2}^*} \nonumber \\&~& +
\varrho_5 e^{\theta_1 + \theta_3 + {\theta_1}^* + {\theta_3}^*}   +
\varrho_6 e^{\theta_1 + \theta_3 + {\theta_2}^* + {\theta_3}^*} \nonumber \\&~& +
\varrho_3^* e^{\theta_2 + \theta_3 + {\theta_1}^* + {\theta_2}^*} +
\varrho_6^* e^{\theta_2 + \theta_3 + {\theta_1}^* + {\theta_3}^*} \nonumber \\&~& +
\varrho_9 e^{\theta_2 + \theta_3 + {\theta_2}^* + {\theta_3}^*}\nonumber \\
{f_6} &=& \Upsilon e^{\theta_1 + {\theta_2} + {\theta_3} +{\theta_1}^* + 
{\theta_2}^* + {\theta_3}^*}\nonumber 
\er

\brs
\begin{aligned}
\theta_1 &= {\bf i}  \frac{{\eta_1}^2}{2} \int D(t) dt +\eta_1 z - {\bf i} \int \lambda dt; \quad
\theta_2 = {\bf i}  \frac{{\eta_2}^2}{2} \int D(t) dt +\eta_2 z - {\bf i} \int \lambda dt  \\
\theta_3 &= {\bf i}  \frac{{\eta_3}^2}{2} \int D(t) dt +\eta_3 z - {\bf i} \int \lambda dt   \quad
\phi^{(3)}  = - {\bf i} \frac{{\zeta_3}^2}{2} \int D(t) dt +{\bf i}  \zeta_3 z - {\bf i}\int \lambda dt \\
\lambda &=  -\gamma R(t) {\sigma_3} \left|\chi^{(3)}(t)\right|^2  
\end{aligned}
\ers
\br
\alpha_1^{(j)}(t) &=& \alpha_{ j1}\sqrt{\frac{D(t)}{\gamma R(t)}};\quad 
\alpha_2^{(j)}(t) = \alpha_{j2}\sqrt{\frac{D(t)}{\gamma R(t)}} \nonumber \\
\alpha_3^{(j)}(t) &=& \alpha_{j3}\sqrt{\frac{D(t)}{\gamma R(t)}}; \quad
\chi^{(3)}(t) = \chi_3\sqrt{\frac{D(t)}{\gamma R(t)}};   \nonumber  
\er

\brs
\begin{aligned}
\beta_{j1} &= (\eta_1-\eta_2)\{\frac{\alpha_1^{(j)}(t) \delta_2^*}{\eta_1 +\eta_1^*} - \frac{\alpha_2^{(j)}(t) \delta_1}{\eta_2 +\eta_1^*}\};\quad
\beta_{j2}  =  (\eta_1-\eta_2)\{\frac{\alpha_1^{(j)}(t) \delta_5}{\eta_1 +\eta_2^*} - \frac{\alpha_2^{(j)}(t) \delta_2}{\eta_2 +\eta_2^*}\} \\
\beta_{j3} &= (\eta_1-\eta_2)\{\frac{\alpha_1^{(j)}(t) \delta_6}{\eta_1 +\eta_3^*} - \frac{\alpha_2^{(j)}(t) \delta_3}{\eta_2 +\eta_3^*}\};\quad
\beta_{j4}  =  (\eta_1-\eta_3)\{\frac{\alpha_1^{(j)}(t) \delta_3^*}{\eta_1 +\eta_1^*} - \frac{\alpha_3^{(j)}(t) \delta_1}{\eta_3 +\eta_1^*}\}\\
\beta_{j5} &= (\eta_1-\eta_3)\{\frac{\alpha_1^{(j)}(t) \delta_6^*}{\eta_1 +\eta_2^*} - \frac{\alpha_3^{(j)}(t) \delta_2}{\eta_3 +\eta_2^*}\};\quad
\beta_{j6}  =  (\eta_1-\eta_3)\{\frac{\alpha_1^{(j)}(t) \delta_9}{\eta_1 +\eta_3^*} - \frac{\alpha_3^{(j)}(t) \delta_3}{\eta_3 +\eta_3^*}\}\\
\beta_{j7} &= (\eta_2-\eta_3)\{\frac{\alpha_2^{(j)}(t)\delta_3^*}{\eta_2 +\eta_1^*} - \frac{\alpha_3^{(j)}(t) \delta_2^*}{\eta_3 +\eta_1^*}\};\quad
\beta_{j8}  =  (\eta_2-\eta_3)\{\frac{\alpha_2^{(j)}(t) \delta_6^*}{\eta_2 +\eta_2^*} - \frac{\alpha_3^{(j)}(t) \delta_5}{\eta_3 +\eta_2^*}\}\\
\beta_{j9} &=  (\eta_2-\eta_3)\{\frac{\alpha_2^{(j)}(t) \delta_9}{\eta_2 +\eta_3^*} - \frac{\alpha_1^{(j)}(t) \delta_6}{\eta_3 +\eta_3^*}\}
\end{aligned}
\ers
\brs
\varrho_1 &=& |\eta_1 -\eta_2|^2 (\frac{\delta_1 \delta_5}{ |\eta_1 +\eta_2^*|^2} - \frac{|\delta_2 |^2}{(\eta_1 +
\eta_1^*)(\eta_2+\eta_2^*)}); \\
\varrho_2 &=& (\frac{\delta_1 \delta_6  (\eta_1 -\eta_2)(\eta_1^* - \eta_3^*)}{ (\eta_1 +\eta_3^*)(\eta_2 +\eta_1^*)} - 
\frac{\delta_3 \delta_2^*  (\eta_1 -\eta_2)(\eta_1^* - \eta_3^*)}{(\eta_1 + \eta_1^*)(\eta_2+\eta_3^*)}) \\
\varrho_3 &=& (\frac{\delta_2 \delta_6  (\eta_1 -\eta_2)(\eta_2^* - \eta_3^*)}{ (\eta_1 +\eta_3^*)(\eta_2 +\eta_2^*)} - 
\frac{\delta_3 \delta_5  (\eta_1 -\eta_2)(\eta_2^* - \eta_3^*)}{(\eta_1 + \eta_2^*)(\eta_2+\eta_3^*)})\\
\varrho_5 &=& |\eta_1 -\eta_3|^2 (\frac{\delta_1 \delta_9}{ |\eta_1 +\eta_3^*|^2} - \frac{|\delta_3|^2}{(\eta_1 +
\eta_1^*)(\eta_3+\eta_3^*)}) \\
\varrho_6 &=& (\frac{\delta_2 \delta_9  (\eta_1 -\eta_3)(\eta_2^* - \eta_3^*) }{ (\eta_1 +\eta_3^*)(\eta_3 +\eta_2^*)} - 
\frac{\delta_3 \delta_6^*  (\eta_1 -\eta_3)(\eta_2^* - \eta_3^*) }{(\eta_1 + \eta_2^*)(\eta_3 + \eta_3^*)}) \\
\varrho_9  &=& |\eta_2 - \eta_3|^2 ( \frac{\delta_5 \delta_9} { |\eta_2 +\eta_3^*|^2} - 
\frac{|\delta_6|^2}{( \eta_3  + \eta_3^* )( \eta_2+\eta_2^* )})
\ers

\brs
\rho_1^{(3)} &=& \varrho_1 \frac{(\xi +i\eta_1)(\xi + i\eta_2)}{(\xi - i\eta_1^*) (\xi - i\eta_2^*)};\qquad
\rho_2^{(3)} = \varrho_2\frac{(\xi +i\eta_1)(\xi + i\eta_2)}{(\xi - i\eta_1^*)(\xi - i\eta_3^*)}\\
\rho_3^{(3)} &=& \varrho_3\frac{(\xi +i\eta_1)(\xi + i\eta_2)}{(\xi - i\eta_2^*)(\xi - i\eta_3^*)};\qquad
\rho_4^{(3)} = \varrho_2^*\frac{(\xi +i\eta_1)(\xi + i\eta_3)}{(\xi - i\eta_1^*)(\xi - i\eta_2^*)}\\
\rho_5^{(3)} &=& \varrho_5\frac{(\xi +i\eta_1)(\xi + i\eta_3)}{(\xi - i\eta_1^*)(\xi - i\eta_3^*)};\qquad
\rho_6^{(3)} = \varrho_6\frac{(\xi +i\eta_1)(\xi + i\eta_3)}{(\xi - i\eta_2^*)(\xi - i\eta_3^*)}\\
\rho_7^{(3)} &=& \varrho_3^*\frac{(\xi +i\eta_2)(\xi + i\eta_3)}{(\xi - i\eta_1^*)(\xi - i\eta_2^*)};\qquad
\rho_8^{(3)} = \varrho_6^*\frac{(\xi +i\eta_2)(\xi + i\eta_3)}{(\xi - i\eta_1^*)(\xi - i\eta_3^*)}\\
\rho_9^{(3)} &=& \varrho_9\frac{(\xi +i\eta_2)(\xi + i\eta_3)}{(\xi - i\eta_2^*)(\xi - i\eta_3^*)}
\ers
\brs
\delta_1 &=& \frac { \gamma R(t)(\displaystyle{\sum_{j=1}^{2} }\sigma_j \left|\alpha_1^{(j)}(t)\right|^2)(\zeta_3 + 
i \eta_1)(\zeta_3- i {\eta_1}^*)}{{(\eta_1 + {\eta_1}^*)}^2(D(t)(\zeta_3 + i \eta_1)(\zeta_3 - i {\eta_1}^*)+
\gamma R(t) \sigma_3 \left|\chi^{(3)}(t)\right|^2)} \\
\delta_2 &=& \frac { \gamma R(t)(\sigma_1 {\alpha_2^{(1)*}(t)}\alpha_1^{(1)}(t)+ \sigma_2 {\alpha_2^{(2)*}(t)} \alpha_1^{(2)}(t))(\zeta_3 + 
i \eta_1)(\zeta_3- i {\eta_2}^*)}{{(\eta_1 + {\eta_2}^*)}^2(D(t)(\zeta_3 + i \eta_1)(\zeta_3 - i {\eta_2}^*)+
\gamma R(t) \sigma_3 \left|\chi^{(3)}(t)\right|^2)}\\
\delta_3 &=& \frac { \gamma R(t)(\sigma_1 {\alpha_1^{(1)}(t)} \alpha_3^{(1)*}(t)+ \sigma_2 {\alpha_1^{(2)}(t)} \alpha_3^{(2)*}(t))
(\zeta_3 - i \eta_3^*)(\zeta_3 + i {\eta_1})}{{(\eta_1 + {\eta_3}^*)}^2(D(t)(\zeta_3 - i \eta_3^*)(\zeta_3 + 
i {\eta_1})+\gamma R(t) \sigma_3 \left|\chi^{(3)}(t)\right|^2)}\\
\delta_5  &=& \frac { \gamma R(t)(\displaystyle{ \sum_{j=1}^{2}} \sigma_j \left|\alpha_2^{(j)}(t)\right|^2)(\zeta_3 + 
i \eta_2)(\zeta_3- i {\eta_2}^*)}{{(\eta_2 + {\eta_2}^*)}^2(D(t)(\zeta_3 + i \eta_2)(\zeta_3 - i {\eta_2}^*)+
\gamma R(t) \sigma_3 \left|\chi^{(3)}(t)\right|^2)} ;\\
\delta_6 &=&  \frac {\gamma R(t)(\sigma_1 {\alpha_3^{(1)*}(t)}\alpha_2^{(1)}(t)+ \sigma_2 {\alpha_3^{(2)*}(t)} \alpha_2^{(2)}(t))(\zeta_3 + 
i \eta_2)(\zeta_3- i {\eta_3}^*)}{{(\eta_2 + {\eta_3}^*)}^2(D(t)(\zeta_3 + i \eta_2)(\zeta_3 - i {\eta_3}^*)+
\gamma R(t) \sigma_3 \left|\chi^{(3)}(t)\right|^2)};\\
\delta_9 &=& \frac { \gamma R(t)(\displaystyle{\sum_{j=1}^{2}} \sigma_j \left|\alpha_3^{(j)}(t)\right|^2)(\zeta_3 + i \eta_3)(\zeta_3- 
i {\eta_3}^*)}{{(\eta_3 + {\eta_3}^*)}^2(D(t)(\zeta_3 + i \eta_3)(\zeta_3 - i {\eta_3}^*)+
\gamma R(t) \sigma_3 \left|\chi^{(3)}(t)\right|^2)} ;
\ers

\brs
\begin{aligned}
\gamma_1^{(3)} &=\delta_1\frac{\xi + i \eta_1}{\xi - i \eta_1^*};
\quad \gamma_2^{(3)} = \delta_2\frac{\xi + i \eta_1}{\xi - i \eta_2^*}; 
\quad  \gamma_3^{(3)}=\delta_3\frac{\xi + i \eta_1}{\xi - i \eta_3^*}; 
\quad \gamma_4^{(3)} = \delta_2^*\frac{\xi + i \eta_2}{\xi - i \eta_1^*}; \\
\gamma_5^{(3)}& = \delta_5\frac{\xi + i \eta_2}{\xi - i \eta_2^*};
\quad\gamma_6^{(3)}=\delta_6\frac{\xi + i \eta_2}{\xi - i \eta_3^*};
\quad \gamma_7^{(3)}=\delta_3^*\frac{\xi + i \eta_3}{\xi - i \eta_1^*}; 
\quad \gamma_8^{(3)} = \delta_6^*\frac{\xi + i \eta_3}{\xi - i \eta_2^*};\\ 
\quad \gamma_9^{(3)}&=\delta_9\frac{\xi + i \eta_3}{\xi - i \eta_3^*}
\end{aligned}
\ers
\brs
\varsigma_{1}^{(j)} & = & \frac{\alpha_1^{(j)}(t) \varrho_3^* (\eta_1 -\eta_2)(\eta_1 -\eta_3)}{ (\eta_1 + \eta_1^*)(\eta_1 + \eta_2^*) } -
                  \frac{\alpha_2^{(j)}(t) \varrho_2^* (\eta_1 -\eta_2)(\eta_2 -\eta_3)}{ (\eta_2 + \eta_1^*)(\eta_2 + \eta_2^*) } \\
                   &+&
                  \frac{\alpha_3^{(j)}(t) \varrho_1 (\eta_1 -\eta_3)(\eta_2 -\eta_3)}{ (\eta_3 + \eta_1^*)(\eta_3 + \eta_2^*) } \\
\varsigma_{2}^{(j)} & = & \frac{\alpha_1^{(j)}(t) \varrho_6^* (\eta_1 -\eta_2)(\eta_1 -\eta_3)}{ (\eta_1 + \eta_1^*)(\eta_1 + \eta_3^*) } -
                  \frac{\alpha_2^{(j)}(t) \varrho_5 (\eta_1 -\eta_2)(\eta_2 -\eta_3)}{ (\eta_2 + \eta_1^*)(\eta_2 + \eta_3^*) } \\
                  &+&
                  \frac{\alpha_3^{(j)}(t) \varrho_2 (\eta_1 -\eta_3)(\eta_2 -\eta_3)}{ (\eta_3 + \eta_1^*)(\eta_3 + \eta_3^*) } \\
\varsigma_{3}^{(j)} & = & \frac{\alpha_1^{(j)}(t) \varrho_9 (\eta_1 -\eta_2)(\eta_1 -\eta_3)}{ (\eta_1 + \eta_2^*)(\eta_1 + \eta_3^*) } -
                  \frac{\alpha_2^{(j)}(t) \varrho_6 (\eta_1 -\eta_2)(\eta_2 -\eta_3)}{ (\eta_2 + \eta_2^*)(\eta_2 + \eta_3^*) } \\
                  &+&
                  \frac{\alpha_3^{(j)}(t) \varrho_3 (\eta_1 -\eta_3)(\eta_2 -\eta_3)}{ (\eta_3 + \eta_2^*)(\eta_3 + \eta_3^*) } \\
\nonumber\\
\ers
\brs
\begin{aligned}
\Upsilon &= \frac{ | (\eta_1 -\eta_2)(\eta_2 -\eta_3)(\eta_1 -\eta_3)|^2}
                           { | (\eta_1 +\eta_2^* )(\eta_1 +\eta_3^*)(\eta_2 +\eta_3^*)|^2}\\
&\times (\delta_1\delta_5\delta_9  +M +M^*\\
&-\frac{\delta_1|\delta_6|^2 |\eta_2 +\eta_3^*|^2}{ (\eta_2 +\eta_2^*)(\eta_3 +\eta_3^*)}
-  \frac{\delta_9|\delta_2|^2 |\eta_1 +\eta_2^*|^2}{ (\eta_1 +\eta_1^*)(\eta_2 +\eta_2^*)} 
-  \frac{\delta_5|\delta_3|^2| \eta_1 +\eta_3^*|^2}{(\eta_1 +\eta_1^*)(\eta_3 +\eta_3^*)}  
           )\\
M &= \delta_2 \delta_6  \delta_3^* \frac{(\eta_2 +\eta_3^*)(\eta_3 +\eta_1^*)(\eta_1 +\eta_2^*)}
{(\eta_1 +\eta_1^*)(\eta_2 +\eta_2^*)(\eta_3 +\eta_3^*)}; \quad 
\nu^{(3)} =  \frac{\Upsilon {(\xi + i \eta_1)(\xi + i \eta_2)(\xi + i \eta_3)}}
{(\xi - i \eta_1^*)(\xi - i \eta_2^*)(\xi - i \eta_3^*)}
\end{aligned}
\ers

\end{document}